\documentclass[preprint,12pt]{elsarticle}
\usepackage{epsfig}
\usepackage{amssymb}
\usepackage{slashed}
\usepackage{amsmath}
\usepackage{cancel}

\journal{Nuclear Physics A}

\newcommand{\gc}{\gamma}

\newcommand{\gk}{\kappa}

\newcommand{\gl}{\lambda}
\newcommand{\gs}{\sigma}

\newcommand{\go}{\omega}

\newcommand{\gL}{\Lambda}

\newcommand{\Y}{{\Upsilon}}
\newcommand{\gC}{\Gamma}

\newcommand{\be}{\begin{equation}}
\newcommand{\ee}{\end{equation}}
\newcommand{\ba}{\begin{eqnarray}}
\newcommand{\ea}{\end{eqnarray}}

\begin{document}

\begin{frontmatter}

\title{Three-Fermion Bound States on the Light Front}

\author{Stefano Mattiello}
\author{Stefan Strauss}
\address{Institut f\"ur Theoretische Physik, Universit\"at Gie{\ss}en, D-35390 Gie{\ss}en, Germany}


\begin{abstract}
We investigate the stability of the relativistic three-fermion system
with a zero-range force in the light front form. In particular, introducing an invariant cut-off, we
study the dependence of the bound state on the coupling strength also for cases where the two-fermion system is unbound.
The relativistic Thomas collapse is discussed by solving the fully coupled integral equation system. 
Furthermore, we explicitly investigate the ground state mass of the three-fermion system and compare to 
previous simplified calculations.
\end{abstract}


\end{frontmatter}

\section{Introduction}
Relativistic constituent quark models in light front dynamics have received
much attention lately~\cite{deAraujo:1995mh}. A major advantage of this
framework is the covariant description
of few-body bound states under kinematical front form boosts~\cite{Keister:1991sb}.
This is a consequence of the stability of the light front Fock state decomposition under boost transformations~\cite{Perry:1990mz}.
Several studies have addressed nucleon properties, such as electromagnetic form factors,
using quarks as the relevant degrees of freedom and Gaussian wave functions
on the light front~\cite{Aznaurian:1982qc,Dziembowski:1987zp,Chung:1991st}.
Other calculations of the proton electric form factors~\cite{deAraujo:1995mh} use nucleon null-plane wave functions obtained from the solution
of the Faddeev equation~\cite{Faddeev:1960su} with a zero-range force~\cite{Frederico:1992uw} acting between the constituent quarks.
In this effective model the spin degrees of freedom and confinement are missing.
Naturally, zero-range interactions provide a simple, but important limiting case
for short range forces and there has been renewed interest in zero-range models with applications in nuclear and atomic physics in the past years~\cite{Fedorov:2001wj,Fedorov:2001vw,Carbonell:2002qs,Yamashita:2008sg}.
The present work is an extension of the developments given in Ref.~\cite{Frederico:1992uw}, where the three-boson bound state problem on the light front was considered. 
Subsequently, the relativistic three-boson problem with zero-range
interactions using light front dynamics has been reformulated
by Carbonell and Karmanov in a covariant way~\cite{Carbonell:2002qs}.
While in the former publication the zero-range interaction was smeared, the
authors of~\cite{Carbonell:2002qs} only assumed the existence of a two-body bound state for the UV regularization without introducing further restrictions in the Faddeev equation itself.
With the latter regularization scheme, which is called $M_{2B}$-regularization in the following, the mass of the three-body bound state may vanish despite the finite value of the mass of the bounded two-body system.
This is the relativistic analog of the so-called Thomas collapse.
Originally, it was pointed out that non-relativistic three-body systems, based on zero-range
forces, experience a Thomas collapse~\cite{Thomas:1935zz}.
Since the inverse of the binding energy $|B_3|^{-1}$ is a measure of the size of the system the three-body state collapses when 
the binding energy is unbounded from below, i.e. $B_3\rightarrow -\infty$. 

A flaw of the $M_{2B}$-regularization scheme, as used in~\cite{Carbonell:2002qs}, is that the explicit dependence of the three-body mass on the coupling constant is not recovered, since the three-body bound state mass is parameterized by the two-body mass.
In order to investigate the three-body bound state equation even if no
two-body bound state exists, the introduction of a different regularization
scheme is necessary. To this end an invariant cut-off $\gL$ as proposed by
Lepage and Brodsky in~\cite{Lepage:1980fj} was introduced into the three-boson
equation in Ref.~\cite{Beyer:2003ag}. More specifically, the requirement is
that the masses of the two- and three-body systems are smaller than the
cut-off, i.e. $M_{20}^2,M_{30}^2<\gL^2$.
It has been shown in Ref.~\cite{Beyer:2003ag} that in the Lepage-Brodsky (LB) regularization scheme the
limit of very large $\Lambda$ coincides with the results of Carbonell and Karmanov where a
comparison can be made. In particular, the relativistic Thomas collapse appears for any value of $\Lambda$.
%

The validity of the bosonic model is evident by the calculation of the proton
electric form factor obtained from the Faddeev wave functions, that reproduces
well the experimental data for low momentum~\cite{deAraujo:1995mh}.
The dominance of this scalar channel for the description of the nucleon static properties has been confirmed by systematic investigations of the relativistic quark coupling effects in the nucleon electromagnetic form factors~\cite{deAraujo:1999hn}.
The scalar coupling between the quark-pair is preferred by neutron data,
independently on the choice of the  momentum wave functions~\cite{deAraujo:2003ke,Suisso:2000hj,deAraujo:2006su}.

However, to achieve a more realistic description of the nucleon the spin
degrees of freedom have to be accounted for. This leads to a coupled system of
integral equations instead of one Faddeev equation.
The first step in this direction has been done again by Karmanov and Carbonell~\cite{Karmanov:2003qk}, where the first results using the $M_{2B}$- regularization for a two-fermion kernel have been reported.
The well-known difficulties to treat the spin on the light front can be
reduced by using kernels factorized relative to initial and final states.
In particular, Karmanov and Carbonell have chosen a simplified kernel with properties similar to the bosonic interaction.
Indeed, the successes of the bosonic model describing the proton properties indicate that this limitation is reasonable. 

In this paper we revisit the interaction given in~\cite{Karmanov:2003qk} using the LB regularization scheme.
In doing so we relax their limitations and investigate 
\begin{itemize}
\item[{\em (i)}] the three-fermion equation even if no two-fermion subsystem
  exist, \item[{\em (ii)}] the direct dependence of the two- and three-body
  bound states on the coupling strength, \item[{\em (iii)}] in contrast
  to~\cite{Karmanov:2003qk}, the full set of coupled integral equations
  and \item[{\em (iv)}] the critical coupling
for the Thomas collapse.
\end{itemize}

This paper is organized as follows.
In the Section~\ref{Sec:Theory} we present the derivation for the two- and three-fermion bound state equation on the light front using a  general zero-range interaction where the kernel is factorized relative to the initial and final states.
We consider a specific interaction kernel corresponding to a $^1S_0$ nucleon in the Section~\ref{Sec:KernelA} and derive the coupled integral Faddeev equations. Our results are compared to the previous calculations of Carbonell and Karmanov and to the three-boson case in the Section~\ref{Sec:Results}. A discussion of the main findings and perspectives is given in the conclusions.

\section{2-Fermion and 3-Fermion bound states for separable kernels}\label{Sec:Theory}
The zero-range two-fermion kernel can be constructed utilizing different spin couplings.
We use kernels being factorized relative to initial and final states, thus having the form
\begin{equation}
{\cal K}_{\gs_1\gs_2}^{\gs'_1\gs'_2}\left(1,2;1',2'\right)=\gl\bar{K}_{\gs_1\gs_2}\left(1,2\right) K^{\gs'_1\gs'_2}\left(1',2'\right),
\end{equation}
with
\be
\bar{K}_{\gs_1\gs_2}\left(1,2\right)\propto\left[\bar{u}_{\gs_1}\left(p_1\right)i\slashed{\go}
  \bar{V} \bar{u}_{\gs_2}\left(p_2\right)\right]
\ee
where $\bar{V}$ indicates the interaction vertex, $\gs_i$ denote the spin
indices of the particles and $u_{\gs_i}$ the corresponding spinors.
The factorization (of the kernel) leads to simplifications that are independent from the specific form of the spin coupling.

\subsection{The two-body problem}

We briefly review the basic ingredients of the two-body equations on
the light front given previously in~\cite{deAraujo:1995mh,Frederico:1992uw,Beyer:2003ag}.
\begin{figure}[t]
\begin{center}
\epsfig{figure=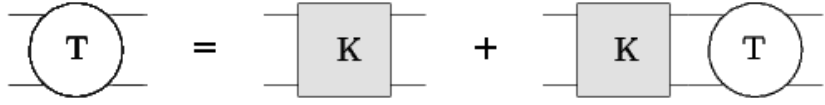,width=0.8\textwidth}
\end{center}
\caption{\label{fig:T2}
  Equation for the two-body $t$-matrix with zero-range interaction.}
\end{figure}
For a zero-range interaction the
equation represented by Fig.~\ref{fig:T2} can be summed leading to a
solution for the two-particle propagator $t(P_2)$, i.e.
\begin{equation}
t(P_2)=\left(i\gl^{-1} - B(P_2)\right)^{-1},
\label{eqn:tau}
\end{equation}
where  $P^\mu_2$ is energy-momentum vector of the two-body system.
The expression for $B(P_2)$ corresponds to a loop diagram
and is given by
\begin{equation}
B(P_2)=-\frac{i}{2(2\pi)^3} \int \frac{dx' d^2 k'_\perp}{x'(1-x')}
\frac{K}{P_2^2-M_{20}^2},
\label{eqn:B}
\end{equation}
where
\begin{eqnarray}\label{def:M_20}
M_{20}^2&=&(\vec k'^{2}_{\perp}+m^2)/x'(1-x'),\\
K&=&\gl^{-1}{\rm Tr}\{{\cal K}\}.
\end{eqnarray}
$M_{20}$ is the mass of the virtual two-particle state and $m$ the mass of the constituent particles.
The integration variables in the loop are $x',\vec k'_\perp$ are the
relative coordinate of the two-body system as defined in Appendix~\ref{Sec:AppK}.
Equation (\ref{eqn:B}) is a generalization of the corresponding equation in the bosonic
case~\cite{deAraujo:1995mh,Frederico:1992uw,Beyer:2003ag},
where the spin structure has been missing. This which is equivalent
to $K=1$ in eq. (\ref{eqn:B}).
The integral (\ref{eqn:B}) has for a contact interaction an UV divergence that has to be regularized. 
If the kernel leads to a logarithmic divergence, like for the scalar bosons, it  can be absorbed in a redefinition of $\gl$ by assuming
that the two-particle propagator $t(P_2)$ has a pole at $P_2^2=M_{2B}^2$.
Then holds the relation 
\begin{equation}
i\gl^{-1}=B(M_{2B}).
\label{eqn:Bound}
\end{equation}
Hence, the renormalization scale of
the propagator is fixed by the mass of the two-particle $M_{2B}$.  
With the $M_{2B}$-regularization the condition (\ref{eqn:Bound}) imposed in the denominator
of eq.~(\ref{eqn:tau}), i.e. $B(M_{2B})-B(P_{2})$,
cancels the UV divergency of $t(P_2)$.
The assumption of the existence of a two-body bound state
regularizes the loop integral completely only for a logarithmic divergence.
For completeness, we remark that, in general, additional conditions have to be
introduced in order to fully regularize the self-energy $B(P_2)$.
Note, that using the $M_{2B}$-regularization scheme the relation between the two-body mass and the coupling constant is not resolved.
The two-particle propagator is the input for the relativistic three-body equations.
In order to investigate the three-body bound state equation also for cases
where no two-body bound state exists another regularization
scheme has to be used. This allows us to find the relation between the coupling constant 
and the two- and three-body masses.

One can introduce an invariant cut-off $\gL$ in the integral (\ref{eqn:B}). The requirement that the 
invariant mass of the two-particle system is smaller than $\gL$ (LB regularization), i.e.
\begin{equation}
M_{20}^2<\Lambda^2,
\label{def:Lambda2}
\end{equation}
makes the integral (\ref{eqn:B}) finite. 
Thus, the integral limits are
\begin{equation}
\int_{0}^{1} dx'\int d^2k'_\perp
\rightarrow 2\pi\int\limits_{x'_{\rm min}}^{x'_{\rm max}}\int\limits_0^{k'_{\rm max}}k'_\perp dk'_\perp,
\label{eqn:Breg}
\end{equation}
where
\begin{eqnarray}
\label{eqn:x1}
x'_{\rm min}&=&\frac{1}{2}\left(1-\sqrt{1-4m^2/\gL^2}\right),\\
\label{eqn:x2}
x'_{\rm max}&=&\frac{1}{2}\left(1+\sqrt{1-4m^2/\gL^2}\right),\\
\label{eqn:km}
k'^2_{\rm max}&=&\gL^2x'(1-x')-m^2.
\end{eqnarray}
Consequently, $t(P_2)\rightarrow t_\Lambda(P_2)$ depends on $\Lambda$ as well.
The LB regularization scheme has been utilized
in~\cite{Beyer:2003ag} and the subsequent
calculations in~\cite{Mattiello:2004rd}.
Accordingly, one can calculate the pole of the two-particle-propagator
determining the bound state mass $M_{2B}$ for any value of the cut-off $\gL$.

\subsection{The three-body problem}\label{Subsection:3B}

\begin{figure}
\begin{center}
    \epsfig{figure=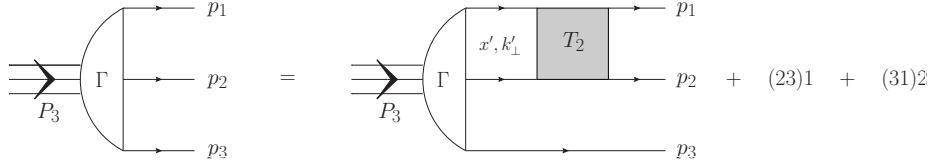,width=0.9\textwidth}
\end{center}
\caption{\label{fig:B3}
  Diagrammatic representation of the Faddeev equation for a zero-range
  interaction, eq. (\ref{eqn:Faddeev-3F}). The gray area indicates the two-body input given in
  Fig.~\protect\ref{fig:T2} and eq.~(\protect\ref{eqn:tau}).}
\end{figure}

Investigating the three-body system, we consider the three-body bound state
wave function $\Psi(1,2,3)$ on the light front, where $1,2,3$ shorthand for $p_1,p_2,p_3$.
The wave function is related to the vertex function $\gC$ by
\be\label{eqn:wave-function}
\Psi(1,2,3)=\frac{\gC(1,2,3)}{{\cal M}^2-M^2_{3B}}\quad\mbox{with}\quad {\cal M}^2=(p_1+p_2+p_3)^2,
\ee
where $M_{3B}$ is the three-body bound state mass.
We expand the vertex function $\gC$ in the standard way as
\begin{equation}
\gC(1,2,3)=\gC_{12}(1,2,3)+\gC_{23}(1,2,3)+\gC_{31}(1,2,3),
\end{equation}
where the Faddeev components $\gC_{ij}(1,2,3)$ satisfy equations involving the interaction between particles $(ij)$ only.
For three fermions the equation for Faddeev component $\gC_{12}(1,2,3)$, by
adding the spin indices, reads
\begin{equation}
\begin{split}
\label{eqn:Faddeev-3F}
\gC_{\gs_1\gs_2\gs_3}^{\gs}(1,2,3)&=\frac{1}{(2\pi)^3}\int\sum_{\gs'_1\gs'_2}{\cal K}_{\gs_1\gs_2}^{\gs'_1\gs'_2}\left(1,2;1',2'\right)\frac{d^2k'_\perp dx'}{2x'(1-x')}\frac{1}{M_{20}^2-P_2^2}\\
&\times\left[\gC_{\gs'_1\gs'_2\gs_3}^{\gs}(1',2',3)+\gC_{\gs'_2\gs_3\gs'_1}^{\gs}(2',3,1')+\gC_{\gs_3\gs'_1\gs'_2}^{\gs}(3,1',2')\right],
\end{split}
\end{equation}
where $\gs$ is the spin projection of the bound state.
A diagrammatic representation of this equation is given in Fig.~\ref{fig:B3}.
In general, as explained in Appendix~\ref{Sec:AppK}, the arguments of $\gC_{\gs_1\gs_2\gs_3}^{\gs}(1,2,3)$ can be
rewritten in variables $\vec k_{i\perp}$ and $x_i$ $(i=1,2,3)$, where the
transversal momenta are $\vec k_{i\perp}=\vec p_{i\perp}-x_i\vec P_{3\perp}$ and the fraction
of the total plus component carried by the single particles is given by
$x_i=p^+_i/P^+_3$. Here $P^\mu_3$ denotes the momentum of the three-body system.
Alternatively, one can rearrange the now defined variables of two particles $i$
and $j$ to the corresponding pair variables $\vec k_\perp$ and $x$ introduced
for the two-body problem.
This is very useful for eq. (\ref{eqn:Faddeev-3F}), where a natural split to a
pair $(ij)$ and to a spectator particle (the third one) emerges.  
The expression for the two-body mass embedded in the three-body system is
\begin{eqnarray}
P_2^2&=&(P_{3} - p_3)^2\nonumber\\
&=&(M_3-p_3^+)\left(M_3-\frac{\vec p_{3\perp}^{~2}+m^2}{p_3^+}\right)
-\vec p_{3\perp}^{~2},
\end{eqnarray}
where the momentum of the three-body system is taken at rest, i.e. $P^\mu_3=(M_3,M_3,0,0)$ in standard light front
coordinates. Using the variables introduced the following relation 
\begin{equation}
P_2^2=(1-x_3)M_3^2-\frac{\vec k_{3\perp}^{~2}+(1-x_3)m^2}{x_3},
\label{eqn:M2}
\end{equation}
holds.

In this work we investigate a bound state with total spin $\gs=1/2$.
Because of the factorized form of the kernels, the Faddeev component
$\gC_{\gs_1\gs_2\gs_3}^{\gs}(1,2,3)$ must be proportional to
$\bar{K}_{\gs_1\gs_2}\left(1,2\right)$ and hence has the form
\be\label{eqn:gamma-G}
\gC_{\gs_1\gs_2\gs_3}^{\gs}(1,2,3)=\bar{K}_{\gs_1\gs_2}\left(1,2\right)G_{\gs_3}^{\gs}(1,2,3).
\ee
Using eq.(\ref{eqn:Faddeev-3F}) one finds the equation for $G_{\gs_3}^{\gs}(1,2,3)$
\begin{equation}
\begin{split}
\label{eqn:Faddeev-G}
G_{\gs_3}^{\gs}(1,2,3)&=\frac{\gl}{(2\pi)^3}\int\sum_{\gs'_1\gs'_2}K_{\gs_1\gs_2}^{\gs'_1\gs'_2}\left(1,2;1',2'\right)\frac{d^2k'_\perp dx'}{2x'(1-x')}\frac{1}{M_{20}^2-P_2^2}\\
&\times\left[\bar{K}_{\gs'_1\gs'_2}(1',2')G_{\gs_3}^{\gs}(1',2',3)+ \bar{K}_{\gs'_2\gs_3}(2',3)G_{\gs'_1}^{\gs}(2',3,1') \right .\\
&+ \left . \bar{K}_{\gs_3\gs'_1}(3,1')G_{\gs'_2}^{\gs}(3,1',2')\right].
\end{split}
\end{equation}
The functions $G_{\gs_3}^{\gs}$  are the components of a $2\times 2$ matrix
labeled by the spin indices.
In order to decompose $G_{\gs_3}^{\gs}$ into basis structures we follow
Refs.~\cite{Karmanov:private,Karmanov:2010ih}, i.e. represent it
as a projection to spin states of four independent elements from the Dirac
space.
Introducing the orthogonal $4\times 4$ matrices $\{S_i\}$ the matrix $G_{\gs_3}^{\gs}$ can be
written as
\be\label{eqn:G-decomp}
G_{\gs_3}^{\gs}(1,2,3)=\sum_{i=1}^4g_i(1,2,3)\bar u_{\gs_3}(p_3)S_i u^\gs(P_3).
\ee
The explicit formulae for the $\{S_i\}$ are given in the Appendix~\ref{Sec:AppS}. Note that the basis elements satisfy the following orthogonality conditions 
\be\label{eqn:S-orth}
{\rm Tr}\left[\bar S_i(\slashed{p}_3+m)S_j(\slashed{P_3}+M_3)\right]=0 \qquad\quad {\rm for }\quad i\neq j,
\ee
with $\bar S_i=\gc_0S_i^\dagger\gc_0$.
The matrices are not normalized to one rather
\be\label{eqn:S-norm}
{\rm Tr}\left[\bar S_i(\slashed{p}_3+m)S_i(\slashed{P_3}+M_3)\right]=N_i=\left\{ \begin{array}{l@{\quad\quad\quad}l}
8x_3k_\perp^{~2}& {\rm for\;} i=1,\\ 8x_3m^2  & {\rm for\;} i=2,\\
8x_3k_\perp^{~2}C_{\rm ps}^2& {\rm for\;} i=3,\\ 8x_3m^2C_{\rm ps}^2  & {\rm for\;} i=4.
\end{array} \right.
\ee
The factors $C_{\rm ps}$ and $C^2_{\rm ps}$ are given by~\cite{Karmanov:1998jp}
\be\label{def:Cps}
C_{\rm ps}=\frac{1}{m^2\go\cdot P_3}\varepsilon^{\mu\nu\rho\gc}p_{1_\mu}p_{2_\nu}P_{3\rho}\go_\gc,\qquad C^2_{\rm ps}=\frac{1}{m^4}\left[k_{3\perp}^2 k^2_\perp-(\vec k_{3\perp}\cdot\vec k_\perp)^2\right],
\ee
where $\go$ is a four-vector with $\go^2=0$ determining the light front orientation by the equation $\go\cdot x=0$.
We use $\go=(1,0,0,-1)$ to recover the standard LF approach.
We keep $\go$ unspecified in the formulae (\ref{def:Cps}) for shortness.
Observe that the matrix $\bar{K}_{\gs_1\gs_2}(1,2)$ in eq.~(\ref{eqn:gamma-G}) is antisymmetric with respect to the permutation $1\leftrightarrow 2$.
The functions $g_{1,2}$ are symmetric, i.e. $g_{1,2}(1,2,3)=g_{1,2}(2,1,3)$.
On the other hand the function $g_{3,4}$ are antisymmetric, i.e. $g_{3,4}(1,2,3)=-g_{3,4}(2,1,3)$, because of the additional antisymmetric factor $C_{\rm ps}$.
Substitution of the decomposition (\ref{eqn:G-decomp}) into eq.~(\ref{eqn:Faddeev-G}) leads to
\begin{eqnarray}
\label{eqn:Faddeev-g}
N_ig_i(1,2,3)&=&\frac{\gl}{(2\pi)^3}\int\frac{d^2k'_\perp dx'}{2x'(1-x')}\frac{1}{M_{20}^2-P_2^2}\\
&\times&\left[KN'_ig_i(1',2',3)+ \sum_{j=1}^4v_{ij}^bg_j(2',3,1')+\sum_{j=1}^4v_{ij}^cg_j(3,1',2')\right].\nonumber
\end{eqnarray}
The factors $N'_{1,2}$ are the same as $N_{1,2}$, whereas in $N'_{3,4}$ the factor $C^2_{\rm ps}$ is replaced by
\begin{eqnarray}
C_{\rm ps}C'_{\rm ps}&=&\left(\frac{1}{m^2\go\cdot P_3}\right)^2\varepsilon^{\mu\nu\rho\gc}p_{1_\mu}p_{2_\nu}P_{3\rho}\go_\gc
\varepsilon^{\bar{\mu}\bar{\nu}\bar{\rho}\bar{\gc}}p'_{1_{\bar{\mu}}}p'_{2_{\bar{\nu}}}
P_{3\bar{\rho}}\go_{\bar{\gc}}\nonumber\\
&=&\frac{1}{m^4}\left((\vec k'_\perp\cdot\vec k_\perp)k^2_{3\perp}-(\vec k_\perp\cdot\vec k_{3\perp})(\vec k'_\perp\cdot\vec k_{3\perp})\right).
\end{eqnarray}
Furthermore the kernels $v_{ij}^b$ and $v_{ij}^c$ are given by
\be
\begin{split}
\label{eqn:vij}
v_{ij}^b&=\sum \bar u_{\gs}(P_3)\bar S_i u^{\gs_3}(p_3)K^{\gs'_1\gs'_2}\left(1',2'\right)\bar{K}_{\gs'_2\gs_3}\left(2',3\right)\bar u_{\gs'_1}(p'_1)S'_{bj} u^{\gs}(P_3),\\
v_{ij}^c&=\sum \bar u_{\gs}(P_3)\bar S_i u^{\gs_3}(p_3)K^{\gs'_1\gs'_2}\left(1',2'\right)\bar{K}_{\gs_3\gs'_1}\left(3,1'\right)\bar u_{\gs'_2}(p'_2)S'_{cj} u^{\gs}(P_3).
\end{split}
\ee
In Appendix~\ref{Sec:AppS} we explicitly explain how the components of $S'_b$
and $S'_c$ are constructed.
The factorization of the contact kernel ensures that $g_3=g_4=0$, as shown in
Ref.~\cite{Karmanov:private}.
Similarly one proofs that $g_1$ and $g_2$ depend on $\vec k_{3\perp}$ and
$x_3$ only, which we rename to $\vec q_\perp$ and $y$ respectively.
Therefore the contribution of $v_{ij}^b$ and $v_{ij}^c$ are equal to each
other.
Additionally, since the first term $g_i$ on the r.h.s. of eq. (\ref{eqn:Faddeev-g})
does not depend on the integration variables, we can
rearrange the integral equation in
terms of the two-particle propagator $t(P_2)$ given in eqs.(\ref{eqn:tau})-(\ref{eqn:B}).
The general form of the integral equations for the three-body bound state reads
\be
\begin{split}
\label{eqn:fad}
g_1(y,\vec q_\perp)&= \frac{t(P_2)}{N_1(2\pi)^3}
\int_{0}^{1} dx'\int d^2k'_\perp\frac{v_{11}g_1(x'(1-y),\vec k'_\perp)+v_{12}g_2(x'(1-y),\vec k'_\perp)}
{(\vec k'_\perp+x'\vec q_\perp)^2+m^2-x'(1-x')P^2_2},\\
g_2(y,\vec q_\perp)&= \frac{t(P_2)}{N_2(2\pi)^3}
\int_{0}^{1} dx'\int d^2k'_\perp
\frac{v_{21}g_1(x'(1-y),\vec k'_\perp)+v_{22}g_2(x'(1-y),\vec k'_\perp)}
{(\vec k'_\perp+x'\vec q_\perp)^2+m^2-x'(1-x')P^2_2}.
\end{split}
\ee

Finally, replacing $x'(1-y)\rightarrow x' $ the equations (\ref{eqn:fad}) is transformed to
\be
\begin{split}
\label{eqn:fad2}
g_1(y,\vec q_\perp)&= \frac{t(P_2)}{N_1(2\pi)^3}
\int_{0}^{1-y} dx'\int d^2k'_\perp
\frac{v_{11}g_1(x',\vec k'_\perp)+v_{12}g_2(x',\vec k'_\perp)}
{P^2_3 -M_{30}^2},\\
g_2(y,\vec q_\perp)&= \frac{t(P_2)}{N_2(2\pi)^3}
\int_{0}^{1-y} dx'\int d^2k'_\perp
\frac{v_{21}g_1(x',\vec k'_\perp)+v_{22}g_2(x',\vec k'_\perp)}
{P^2_3 -M_{30}^2},
\end{split}
\ee
where $M^2_{30}$ is the mass square of the virtual three-particle state in the rest system, i.e.
\begin{equation}\label{def:M_30}
M_{30}^2=\frac{\vec k'^{2}_\perp+m^2}{x'}
+\frac{\vec q^{~2}_\perp+m^2}{y}
+\frac{(\vec k'+\vec q)^2_\perp+m^2}{1-x'-y}.
\end{equation}
In order to consistently perform the LB regularization in the two- as well in
the three-fermion equations we also constrain the mass square of the virtual
three-particle state by
\begin{equation}\label{eqn:thetaM30}
\int_{0}^{1-y} dx'\int d^2 k'_\perp
\rightarrow \int_{0}^{1-y} dx'
\int d^2 k'_\perp \theta(M^2_{30}-\Lambda^2).
\end{equation}

After fixing an explicit interaction kernel and spin structure the aim is to
solve the eigenvalue problem posed by eqs.~(\ref{eqn:fad2}) in order to obtain
the mass $M_3^2=P_3^2$ of the three-fermion system as a function of the the coupling constant $\gl$.

\section{The $^1S_0$ Model}\label{Sec:KernelA} 
In this section we investigate a particular kernel which leads to significant
simplifications in the three-body equations (\ref{eqn:fad2}).
The spin structure of this kernel is given by
\be
\begin{split}\label{eqn:KernA}
\bar{K}_{\gs_1\gs_2}\left(1,2\right)&=
\frac{m}{\go\cdot(p_1+p_2)}
\left[\bar{u}_{\gs_1}\left(p_1\right)i\slashed{\go} \gc_5U_{\mathrm{c}}\bar{u}_{\gs_2}\left(p_2\right)\right],\\
K_{\gs'_1\gs'_2}\left(1',2'\right)&=
\frac{m}{\go\cdot(p_1+p_2)}
\left[u^{\gs'_2}\left(p'_{2}\right)U_{\mathrm{c}}i\slashed{\go}\gc_5u^{\gs'_1}\left(p'_{1}\right)\right],
\end{split}
\ee
where $U_{\mathrm{c}}=i\gc^0\gc^2$ denotes the charge conjugation operator.
In the nonrelatististic limit the kernel reads
\ba
{\cal K}_{\gs_1\gs_2}^{\gs'_1\gs'_2}\left(1,2;1',2'\right)&\propto&
\frac{\gl}{2} \left(w_{\gs_1}^{*}i\gs_yw_{\gs_2}^{*}\right)
\left(w_{\gs'_2}i\gs_yw_{\gs'_1}\right)\\
&=&\gl C^{00}_{(1/2)\gs_1(1/2)\gs_2}C^{00}_{(1/2)\gs'_1(1/2)\gs'_2}\left(w_{\gs_1}^{*}i\gs_yw_{\gs_2}^{*}\right)\left(w_{\gs'_2}i\gs_yw_{\gs'_1}\right)\nonumber,
\label{eqn:KernA-nonrel}
\ea
where $w_\gs$ represent the two-component (Pauli) spinors and
$C^{00}_{(1/2)\gs_1(1/2)\gs_2}$ are the Clebsh-Gordon coefficients~\cite{Carbonell:1998rj}.
Therefore, this kernel corresponds to an interaction in the $^1\mbox{S}_0$ state.\\
For this interaction we find
\ba
K&=&\gl^{-1}{\rm Tr}\{{\cal K}\}\nonumber\\
&=&\frac{m^2}{(\go\cdot P_2)^2}\sum_{\gs'_1\gs'_2}
\left[u^{\gs'_2}\left(p'_{2}\right)U_{\mathrm{c}}i\slashed{\go}\gc_5u^{\gs'_1}\left(p'_{1}\right)\right]
\left[\bar{u}_{\gs'_1}\left(p'_1\right)i\slashed{\go}\gc_5U_{\mathrm{c}}\bar{u}_{\gs'_2}\left(p'_2\right)\right]\nonumber\\
&=&8\gl m^2x(1-x).
\ea
From equation (\ref{eqn:B}) the loop integral $B_2(P_2)$ reads
\be
B(P_2)=-i\frac{4m^2}{(2\pi)^3} \int dx' d^2k'_\perp
\frac{1}{P_2^2-M_{20}^2}.
\label{eqn:B-KA}
\ee
Using the $M_{2B}$ regularization 
one finds
\be
t(P_2)=\frac{48\pi^2}{4m^2}\gk(P_2,M_{2B})
\ee
where
\ba
\gk(P_2,M_{2B})&=&\left[\frac{3\Y^2(P_2)+1}{\Y^3(P_2)}\arctan\Y(P_2)
-\frac{1}{\Y^2(P_2)} \right . \nonumber\\
&-& \left . \frac{3\Y^2(M_{2B})+1}{\Y^3(M_{2B})}\arctan\Y(M_{2B})
+\frac{1}{\Y^2(M_{2B})}\right]^{-1}
\ea
with
\be
\Y(P_2)=\left\{ \begin{array}{l@{\quad\quad\quad}l}
\frac{P_2}{\sqrt{4m^2-P^2_2}}& {\rm for\;} 0\leq P^2_{2}\leq 4m^2,\\ \frac{\sqrt{-P^2_2}}{\sqrt{4m^2-P^2_2}}  & {\rm for\;} P^2_2<0
\end{array} \right..
\ee

Introducing the invariant cut-off as in eq.(\ref{def:Lambda2}) the loop
integral can be directly evaluated using the limits of the integration given in
eqs.(\ref{eqn:x1})-(\ref{eqn:km}).
In this way, for any value of the cut-off and
the coupling we can calculate the pole of the two-fermion propagator.\\
Now we turn to the three-fermion case.
Calculating the $2\times 2$ matrix for the kernel (\ref{eqn:KernA}) using eq.(\ref{eqn:vij}) yields
\be
\begin{split}
v_{11}&=32m^2yx'^2\left(\vec q_\perp^{2}+\vec k'_\perp\cdot\vec q_\perp\right),\\
v_{12}&=0,\\
v_{21}&=32m^4yx'\left(1-2x'\right),\\
v_{11}&=-32m^4yx'\left(1-x'\right).
\end{split}
\ee
Shifting $x'(1-y)\rightarrow x'$ and using $z= x'/(1-y)$ we find the following
set of equations
\ba
\label{eqn:fad-KA1-final}
g_1(y,\vec q_\perp) &=& \frac{4m^2t(P_2)}{(2\pi)^3}
\int_{0}^{1-y} \frac{dx'}{x'(1-y-x')}\int d^2k'_\perp\theta(M^2_{30}-\Lambda^2)\nonumber\\
&&\frac{(1+\vec k'_\perp\cdot\vec q_\perp/\vec q_\perp^{~2})z^2}
{P^2_3 -M_{30}^2}\;g_1(x',\vec k'_\perp),\\
g_2(y,\vec q_\perp) &=& \frac{4m^2t(P_2)}{(2\pi)^3}
\int_{0}^{1-y} \frac{dx'}{x'(1-y-x')}\int d^2k'_\perp\theta(M^2_{30}-\Lambda^2)\nonumber\\
&&
\frac{z(1-2z)g_1(x',\vec k'_\perp)-z(1-z)g_2(x',\vec k'_\perp)}
{P^2_3 -M_{30}^2}.
\label{eqn:fad-KA2-final}
\ea

The scalar function $g_2$ enters in the second equation only. The first equation differs from the three-boson equation by the factor $(1+\vec k'_\perp\cdot\vec q_\perp/\vec q_\perp^{~2})z^2$.
We point out that the complete system of equations must be solved, not only the first equation as in Ref.~\cite{Karmanov:2003qk}.


\section{Results}
\label{Sec:Results}
The solution of the equations
(\ref{eqn:fad-KA1-final})-(\ref{eqn:fad-KA2-final}) allows us to determine the three-fermion mass.
In Ref.~\cite{Karmanov:2003qk} eq.~(\ref{eqn:fad-KA1-final}) is solved using the $M_{2B}$ regularization.
Instead we consider the LB regularized version of~(\ref{eqn:fad-KA1-final})-(\ref{eqn:fad-KA2-final}). 
In order to compare our results with those of~\cite{Karmanov:2003qk} we set the cut-off to
a large but finite value, i.e. $\gL=10^{15}m$, see Ref.~\cite{Beyer:2003ag}. For the bosonic system it was shown by one of the authors that for large invariant cut-off the
resulting function $M_{3B}(M_{2B})$ coincides with the analogous results in
$M_{2B}$ regularization~\cite{Beyer:2003ag,Mattiello:2004rd}.
Therefore, we expect this feature as well in the fermionic case.


\begin{figure}[t]
\begin{center}
    \epsfig{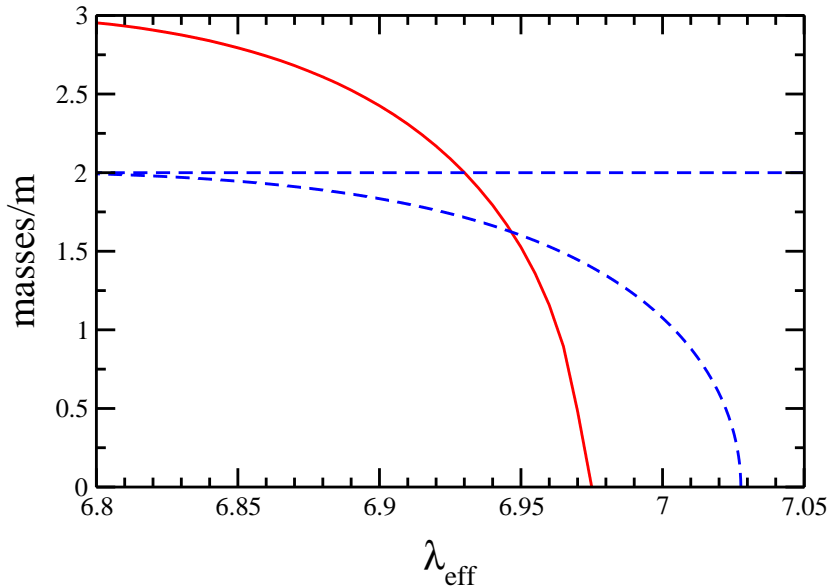}
\end{center}
\caption{\label{fig:m23L}
  The solution of the two- (dashed) and three-fermion (solid) bound state
  equations as a function of the effective coupling $\lambda_{\rm eff}=\lambda m^2$.  Horizontal
  dashed line shows the two-fermion break-up.}
\end{figure}

Fig.~\ref{fig:m23L} shows the masses $M_{2B}$ of the two-fermion bound state
and $M_{3B}$ of the three-fermion bound
state as a function of the effective coupling strength $\lambda_{\rm
  eff}=\lambda m^2$. All masses $M_{2B}$ and $M_{3B}$ are given in units of the elementary fermion mass $m$.

Note the following features:
\begin{itemize}
\item[{\em (i)}] The two- as well as the three-fermion mass are monotonic function of the coupling. In particular, for the critical coupling $\lambda^{\rm c}_{\rm eff}\approx 6.972$ the three-fermion bound
state mass vanishes. This is the relativistic analog of the Thomas
collapse~\cite{Thomas:1935zz} and is known to occur for the three-boson bound state with a similar zero-range interaction, see e.g. \cite{Beyer:2003ag,Mattiello:2004rd}. This critical coupling corresponds to a critical value of two-fermion mass $M^{\rm c}_{2B}\approx 1.42m$.
In comparison, the critical two-fermion mass $M^{\rm c}_{2B}\approx 1.35m$ calculated in~\cite{Karmanov:2003qk} is notable lower.

\item[{\em (ii)}] As $M_{2B}\rightarrow 2m$, the three-fermion bound state still exists.
This may give rise to the Efimov effect~\cite{Efimov:1970zz,Braaten:2004rn}, an interesting issue for further investigations.

\item[{\em (iii)}] There is a region of parameter space where both $M_{2B}$ and
$M_{3B}$ exist. Only in this case it is possible to plot
$M_{3B}$ against $M_{2B}$. 
\end{itemize}

\begin{figure}[t]
\begin{center}
    \epsfig{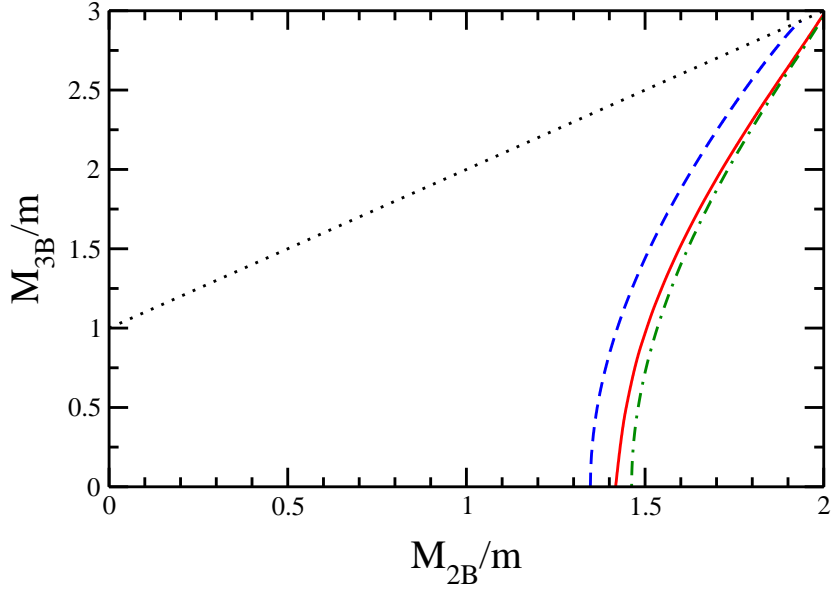}
\end{center}
\caption{\label{fig:KA-M23}
  Three-fermion bound state as a function of $M_{2B}$ 
  (solid line) in comparison to the three-boson bound state (dash-dotted line) and to three-fermion bound state calculated by solving equation  (\ref{eqn:fad-KA1-final}) alone, with $\Lambda\rightarrow\infty$ (dashed line). To completeness, the threshold for the three-body mass is indicated (dotted line).}
\end{figure}

We focus on the investigation in the region of the coupling strength $\gl$, where the two-fermion bound state exists.
Therefore, by comparison to the results of~\cite{Karmanov:2003qk}, we can estimate the role of the coupled structure of (\ref{eqn:fad-KA1-final})-(\ref{eqn:fad-KA2-final}).
In Fig.~\ref{fig:KA-M23} the mass $M_{3B}$ is plotted against $M_{2B}$. 
We have also included the solid line for fermions and the dash-dotted line for bosons.
For comparison to the calculation of~\cite{Karmanov:2003qk} we independently solve, using the LB regularization scheme, eq. (\ref{eqn:fad-KA1-final}) solely and present the results in Fig.~\ref{fig:KA-M23} as dashed line.
Furthermore, the dotted line indicates the threshold for the three-body mass.

The resulting functions $M_{3B}(M_{2B})$ for the bosonic and for the fermionic cases have the same qualitative behavior. 
By decreasing $M_{2B}$ the three-body mass decreases very quickly and vanishes at a critical values $M^{\rm c}_{2B}$ of the two body mass.
The value of the critical mass for the bosonic system is $M^{\rm
  c}_{2B}\approx 1.45m$ and greater than the corresponding values for the
fermionic cases, both for the coupled equation system and for the solution of
the first equation only.

It is interesting to see why there is this notable discrepancy between the
whole calculation and the result obtained by solving (\ref{eqn:fad-KA1-final}) alone.

The latter method automatically excludes the trivial solution $g_1\equiv 0$ for
eq. (\ref{eqn:fad-KA1-final}).
However, this solution is physically non-trivial in the coupled system.

In more detail, the system of integral equations~(\ref{eqn:fad-KA1-final})-(\ref{eqn:fad-KA2-final}) can be written as
\be
(I-V)\left(
\begin{array}{c}
g_1\\
g_2\end{array}
\right)=0
\label{eqn:evv}
\ee
where the operator $V$ collects all contributions from $v_{ij}$ and $I$ is the identity operator.
 
To achieve a numerical solution we have discretized the equations
(\ref{eqn:fad-KA1-final})-(\ref{eqn:fad-KA2-final}) using a momentum space
grid ${\tau_k}$, with $k=1,\cdots,n$, where $n$ is the number of the
lattice points. Each of the scalar function $g_i$,  $i=1,2$ becomes a vector
$\tilde g_1$ of dimension $n$.
The operator eigenvalue problem~(\ref{eqn:evv}) is approximated by the corresponding matrix problem
\be\label{eqn:Wg}
W\;\tilde g=0\quad{\rm with} \quad \tilde g=\left(
\begin{array}{c}
\tilde g_1\\
\tilde g_2\end{array}
\right),
\ee
where
\be
W=\left(
\begin{array}{cc}
W_{11}&W_{12}\\
W_{21}&W_{22}\end{array}
\right).
\ee
All possible solutions fulfill of course the condition $\det(W)=0$.
For the $^1S_0$ model~(\ref{eqn:KernA}) one has $W_{12}=0$ and the solvability condition reads
\be
\det(W)=\det(W_{11})\cdot \det(W_{22})=0.
\ee
Evidently, a solution of the whole system satisfies $\det(W_{11})=0$ {\em or} $\det(W_{22})=0$, whereas the solutions of eq.(\ref{eqn:fad-KA1-final}) satisfies $\det(W_{11})=0$ only.
Hence, solving solely eq.(\ref{eqn:fad-KA1-final}) the physical solutions
corresponding to the case $g_1\equiv 0$ can be lost.

\begin{figure}[t]
\begin{center}
    \epsfig{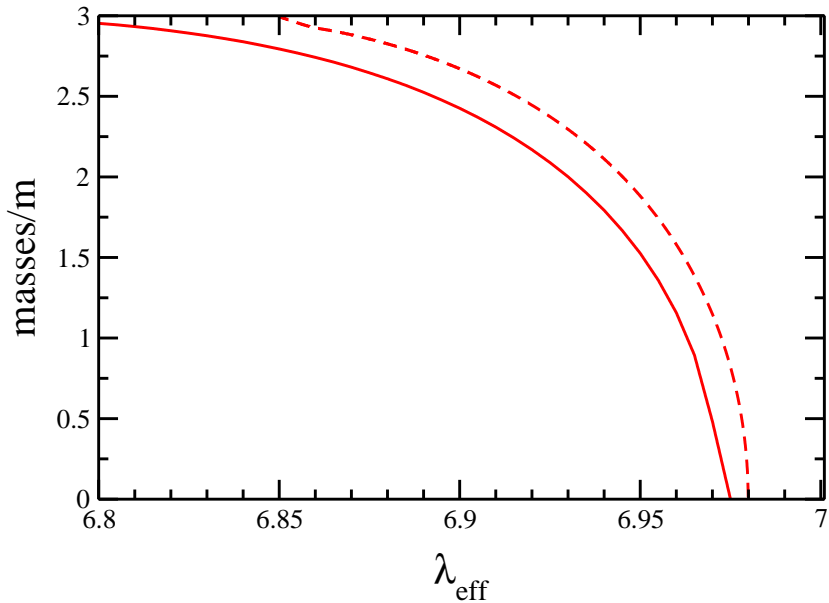}
\end{center}\caption{Mass of the three-fermion ground state (solid line) and of the first three-fermion excited state (dashed line) calculated by solving the equation system (\ref{eqn:fad-KA1-final})-(\ref{eqn:fad-KA2-final}) as a function of the effective strength $\lambda_{\rm eff}$.
\label{fig:M3N0N1}}
\end{figure}

\begin{figure}[t]
\begin{center}
    \epsfig{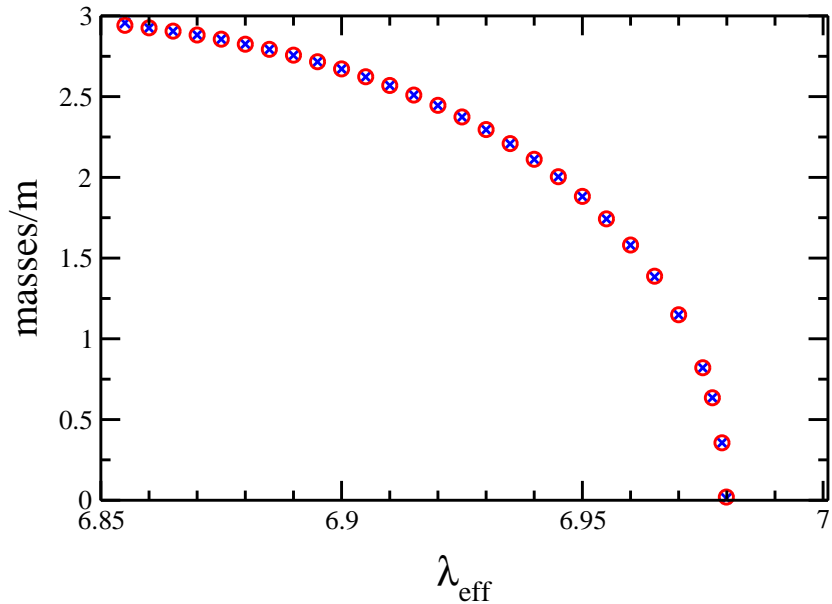}
\end{center}\caption{Mass of the three-fermion bound state  calculated by solving eq.(\ref{eqn:fad-KA1-final}) only (crosses) and mass of the first three-fermion excited state calculated by solving the set of equations~(\ref{eqn:fad-KA1-final})-(\ref{eqn:fad-KA2-final}) (circles) as a function of the effective strength $\lambda_{\rm eff}$.
\label{fig:M3-Lcomp}
}
\end{figure}

Because for arbitrary coupling strength $\gl$ there exist more solutions to (\ref{eqn:Wg}), we consider the two smallest eigenvalues. 
In Fig.~\ref{fig:M3N0N1} the mass of the ground state (lowest eigenstate) and the first excited state (next-to-lowest eigenstate) depending on effective coupling strength $\lambda_{\rm eff}$ are plotted as solid and dashed line, respectively.
The ground state coincides with the solution shown in Fig.~\ref{fig:m23L}.
For the first excited state we find that the Thomas collapse occurs at
$\gl_{\rm eff}\approx 6.98$ that corresponds to a critical two-fermion mass $M^{\rm c}_{2B}\approx 1.35m$ which is in agreement with the critical value obtained by solving eq.(\ref{eqn:fad-KA1-final}) only.
We therefore suspect that the solution given in Ref.~\cite{Karmanov:2003qk} describes the first excited state. This observation is 
confirmed in Fig.~\ref{fig:M3-Lcomp} where the bound state mass solving only eq.~(\ref{eqn:fad-KA1-final}) (crosses) is compared with the solution for the first excited state (circles) obtained by solving the coupled equation system (\ref{eqn:fad-KA1-final})-(\ref{eqn:fad-KA2-final}) for different values of the effective coupling strength $\gl_{\rm eff}$.
The two masses agree well for all $\gl_{\rm eff}$ but small derivations are noticeable close to the breakup, i.e. for $\gl_{\rm eff} < 6.86$.
In other words the solution of eq.(\ref{eqn:fad-KA1-final}) describes three-fermion bound states, but the ground state is missing. The
true ground state is only obtained in the solution of the coupled system of
integral equations.

\section{Conclusions}\label{Sec:Conclusion}
We have derived and solved the coupled equations of a relativistic three
fermion system  subject to an effective scalar zero-rage interaction in the $^1S_0$ channel.
In the computation of the three- and two-fermion bound states as function of
the coupling constant $\gl$ we have introduced an invariant cut-off $\Lambda$
(LB regularization scheme) and solved the equations for large $\Lambda$.
Additionally, we have investigated the solution of eq.~(\ref{eqn:fad-KA1-final}) alone with the same regularization scheme.
Thereby the results of Ref.~\cite{Karmanov:2003qk} were confirmed.
Analogous to the three-boson system we find the relativistic Thomas collapse in both calculations.
In particular, solving the coupled equations we find that the two-fermion critical mass $M^{\rm c}_{2B}\approx 1.42m$ is different from the value given in~\cite{Karmanov:2003qk}.
This is explained by the fact that in the computation of Carbonell/Karmanov
the lowest three-fermion mass is missing. Instead of the ground state their computed lowest mass seems to agree with the mass of the first excited state. 
Taking in account the full equation system, we have obtained a
critical two-body mass different from the one in the literature and more close
to the critical value in the bosonic case (see Fig.~\ref{fig:KA-M23}).

In particular, our results provide evidence we that the bosonic or fermionic nature of the constituents does not play a very important role in this model.
Consequently, using the bosonic-type Faddeev equations to describe the
isolated nucleon~\cite{Beyer:2003ag} or embedded in a quark matter
medium~\cite{Beyer:2001bc,Strauss:2009uj} can be justified {\em a posteriori}.

Nevertheless, an implementation of the spin degrees of freedom
is desirable and a first consistent step in this direction is studied here.
Further investigations of the three-fermion correlations in hot and
dense quark matter following~\cite{Strauss:2009uj} can be performed
using the interaction kernel (\ref{eqn:KernA}).

The theoretical methods presented here are model independent and allow
the investigation of the nucleon with more complicated kernels, such as the
one corresponding to an interaction in the $P$-wave state
with $S=1$~\cite{Karmanov:private}.

Acknowledgment: 
The authors are grateful to V.A. Karmanov for the providing his theoretical
notes~\cite{Karmanov:private}. This work is supported by Deutsche
Forschungsgemeinschaft and by the Helmholtz International Center for FAIR
within the LOEWE program of the State of Hesse.

\appendix

\section{Kinematics}\label{Sec:AppK}

Let $a^\mu=(a^0,a^1,a^2,a^3)$ be a generic four vector.  The light front
dynamics is given by the transformation
\be\label{+-def}
a^\mu = \left(a^+,a^-,a^1,a^2\right) =
\left(a^+,a^-,\vec{a}_\perp\right) ,
\ee
where 
\be
a^\pm\equiv a^0\pm a^3.
\ee

The covariant components follow from $a_\mu=g_{\mu\nu} a^\nu$, where the metric
tensor is given as,
\begin{equation}
g_{\mu\nu}=
\left(
\begin{array}{cccc}
{0}&{\frac{1}{2}}&{0}&{0}\\
{\frac{1}{2}}&{0}&{0}&{0}\\
{0}&{0}&{-1}&{0}\\
{0}&{0}&{0}&{-1}\end{array}
\right),
\; g^{\mu\nu}=
\left(
\begin{array}{cccc}
{0}&{2}&{0}&{0}\\
{2}&{0}&{0}&{0}\\
{0}&{0}&{-1}&{0}\\
{0}&{0}&{0}&{-1}\end{array}
\right).
\end{equation}
The mass shell condition for a particle of mass $m$ and momentum $p$
reads
\begin{equation}
      p^+p^- = {\vec p_\perp}^2 +m^2.
\label{eqn:onshell}
\end{equation}

For the description of the kinematics of two- and three-particle systems it is
convenient to introduce the total momentum of the particles and some internal
variables.
For two particles with momenta $p_1$ and $p_2$ and masses $m_1$ and $m_2$
respectively we define
\ba
P_2&=&p_1+p_2, \\
x&=&p_1^+/P_2^+, \quad 1-x=p_2^+/P_2^+,\\
k&=&(1-x)p_1-xp_2.
\ea
The invariant energy
square, $s_{12}$, i.e. the mass of the virtual two-body state, is given by
\ba
s_{12}\equiv M_{20}^2&=&P_2^+ P_2^- -\vec P_{2\perp}^2=\sum\limits_{i=1}^{2}\frac{\vec p_{i\perp}^2+m_i^2}{k_i^+}P_2^+-\vec P_{2\perp}^2\\
&=& \frac{\vec p_{1\perp}^2+m_1^2}{x}+\frac{\vec p_{2\perp}^2+m_2^2}{1-x}-\vec
P_{2\perp}^2\\
&=&\frac{m_2^2}{1-x} +\frac{m_1^2}{x}+\frac{\vec k_\perp^2}{x(1-x)}
\ea
Choosing $\vec P_{2\perp}=0$ and for equal masses, i.e. $m_1=m_2\equiv m$ we
recover the expression in eq.(\ref{def:M_20}).

For three particles with momenta $p_1$, $p_2$ and $p_3$ and masses $m_1$,
$m_2$ and $m_3$
we analogously introduce the plus momentum fractions and the total momentum
as
\ba\label{eqn:kin3b-1}
P_3&=&p_1+p_2+p_3, \\
x_i&=&p_i^+/P_3^+, \quad \sum\limits_{i=1}^{3}x_i=1.\\
\ea
In general one can introduce Jacobi coordinates to treat the kinematics of
the three-body system~\cite{Bakker:1979eg}.
However, because of the structure of the Faddeev-equation, where a spectator
particle and a pair of two particles emerge, we prefer to express
the kinematics using the variable of the spectator particle -say the third
particle - and the one of the pair $(12)$, where kinematics has been discussed
above.
Note that the momentum fraction of the three-body system $x_i$ and the
variable $x$ do not coincide but are related by
\ba\label{eqn:permut-x}
x_1&=&x(1-x_3),\\
x_2&=&(1-x)(1-x_3).
\ea
The additional constraint used in this paper, namely that the three-body system is at
rest, allows us to connect the single transverse momenta of the
particle in the pair with the relative transverse momentum $\vec k_\perp$ by
\ba
\vec p_{1\perp}&=&-\vec k_\perp-x\vec p_{3\perp},\\
\vec p_{2\perp}&=&\vec k_\perp-(1-x)\vec p_{3\perp}.
\ea
The invariant energy
square, $s_{123}$, i.e. the mass of the virtual three-body state, can be
written as
\ba
s_{123}\equiv M_{30}^2&=&P_3^+ P_3^- -\vec P_{3\perp}^2=\sum\limits_{i=1}^{3}\frac{\vec p_{i\perp}^2+m_i^2}{k_i^+}P_3^+-\vec P_{3\perp}^2\\
&=& \sum\limits_{i=1}^{3}\frac{\vec p_{i\perp}^2+m_i^2}{x_i^+}-\vec P_{3\perp}^2.
\ea
Demanding $\vec P_{3\perp}=0$ and for equal masses, i.e. $m_1=m_2=m_3\equiv m$ we
recover, with the necessary variable identification, eq.(\ref{def:M_30}).

\section{Orthogonal basis}\label{Sec:AppS}
In this section we define the orthogonal basis $\{S_i\}$ as given in
Refs.~\cite{Karmanov:private,Karmanov:2010ih} used for the decomposition of the $2\times 2$-matrix $G_{\gs_3}^{\gs}$
in eq. (\ref{eqn:G-decomp}).
We choose the form
\ba
\label{def:S1}
S_1&=&\left[2x_3-(m+x_3M_3)\frac{\slashed{\go}}{\go\cdot P_3}\right],\\
\label{def:S2}
S_2&=&m\frac{\slashed{\go}}{\go\cdot P_3},\\
\label{def:S3}
S_3&=&\left[2x_3-(m-x_3M_3)\frac{\slashed{\go}}{\go\cdot P_3}\right]iC_{\rm ps}\gc_5,\\
\label{def:S4}
S_4&=&m\frac{\slashed{\go}}{\go\cdot P_3}iC_{\rm ps}\gc_5,
\ea
where the factor $C_{\rm ps}$ is given in eq.(\ref{def:Cps}).\\
Furthermore, the matrices $\bar S_i=\gc_0S_i^\dagger\gc_0$ explicitly read
\ba
\bar S_1&=&\left[2x_3-(m+x_3M_3)\frac{\slashed{\go}}{\go\cdot P_3}\right],\\
\bar S_2&=&m\frac{\slashed{\go}}{\go\cdot P_3},\\
\bar S_3&=&iC_{\rm ps}\gc_5\left[2x_3-(m-x_3M_3)\frac{\slashed{\go}}{\go\cdot P_3}\right],\\
\bar S_4&=&iC_{\rm ps}\gc_5m\frac{\slashed{\go}}{\go\cdot P_3}.
\ea
Note that $\bar S_{1,2}=S_{1,2}$.
These matrices satisfy the orthogonality (\ref{eqn:S-orth}) and normalization
conditions (\ref{eqn:S-norm}).

The permutated structures $S'_{b}$ and $S'_{c}$ are derived from eq.~(\ref{def:S1}-\ref{def:S4}).
$S'_{b}$ is constructed using  the permutation
$(1'\rightarrow 2'\rightarrow 3\rightarrow 1')$, e.g. for $S'_{b1}$ one
performs the substitution $x_3\rightarrow x'_1=(1-x)(1-x_3)$ in
eq.~(\ref{def:S1}), while for $S'_{b3}$ and $S'_{b4}$ one additionally has to
replace $C_{\rm ps}$ by
\be
C'_{\rm ps}=\frac{1}{m^2\go\cdot P_3}\varepsilon^{\mu\nu\rho\gc}p'_{1_\mu}p'_{2_\nu}P_{3\rho}\go_\gc\rightarrow\frac{1}{m^2\go\cdot P_3}\varepsilon^{\mu\nu\rho\gc}p'_{2_\mu}p_{3_\nu}P_{3\rho}\go_\gc
\ee
in eq.~(\ref{def:S3}) and (\ref{def:S4}) respectively. 
Analogously, $S'_{c}$ is derived from $S$ by the permutation
$(1'\rightarrow 3\rightarrow 2'\rightarrow 1')$, e.g. for $S_{c1}$ the
substitution $x_3\rightarrow x'_2=x(1-x_3)$  has to be used in
eq.~(\ref{def:S1}), while for $S'_{c3}$ and $S'_{c4}$ one has to use
\be
C'_{\rm ps}=\frac{1}{m^2\go\cdot P_3}\varepsilon^{\mu\nu\rho\gc}p'_{1_\mu}p'_{2_\nu}P_{3\rho}\go_\gc\rightarrow\frac{1}{m^2\go\cdot P_3}\varepsilon^{\mu\nu\rho\gc}p_{3_\mu}p_{1_\nu}P_{3\rho}\go_\gc
\ee
instead of $C_{\rm ps}$ in eq.~(\ref{def:S3}) and (\ref{def:S4}) respectively.

\bibliographystyle{elsarticle-num-names}
\bibliography{literature}

\end{document}